%%%%%%%%%%%%%%%%%%%%%%%%%%%%%%%%%%%%%%%%%%%%%%%%%%%%%%%%%%%%%%%%%%%%
%  TeX Definitions                                                 %
%%%%%%%%%%%%%%%%%%%%%%%%%%%%%%%%%%%%%%%%%%%%%%%%%%%%%%%%%%%%%%%%%%%%

\newif\iffigs\figstrue
% Uncomment the next line if you do not want the figures
%\figsfalse

%
% the following is to use blackboard bold fonts --
\let\useblackboard=\iftrue
%
% activate this if you don't have them.
%\let\useblackboard=\iffalse
%
% You might also need to remove this line.
\newfam\black

\input harvmac.tex

\iffigs
  \input epsf
\else
  \message{No figures will be included.  See TeX file for more
information.}
\fi

\def\Title#1#2{\rightline{#1}
\ifx\answ\bigans\nopagenumbers\pageno0\vskip1in%
\baselineskip 15pt plus 1pt minus 1pt
\else%\special{papersize=11in,8.5in}%
\def\listrefs{\footatend\vskip 1in\immediate\closeout\rfile\writestoppt
\baselineskip=14pt\centerline{{\bf References}}\bigskip{\frenchspacing%
\parindent=20pt\escapechar=` \input
refs.tmp\vfill\eject}\nonfrenchspacing}
\pageno1\vskip.8in\fi \centerline{\titlefont #2}\vskip .5in}

\ifx\answ\bigans\def\tcbreak#1{}\else\def\tcbreak#1{\cr&{#1}}\fi
\useblackboard
\message{If you do not have msbm (blackboard bold) fonts,}
\message{change the option at the top of the tex file.}
\font\blackboard=msbm10 %scaled \magstep1
\font\blackboards=msbm7
\font\blackboardss=msbm5
%\newfam\black
\textfont\black=\blackboard
\scriptfont\black=\blackboards
\scriptscriptfont\black=\blackboardss
\def\Bbb#1{{\fam\black\relax#1}}
\else
\def\Bbb#1{{\bf #1}}
\fi
% *************************************
%
\def\yboxit#1#2{\vbox{\hrule height #1 \hbox{\vrule width #1
\vbox{#2}\vrule width #1 }\hrule height #1 }}
\def\fillbox#1{\hbox to #1{\vbox to #1{\vfil}\hfil}}
\def\ybox{{\lower 1.3pt \yboxit{0.4pt}{\fillbox{8pt}}\hskip-0.2pt}}
\def\np#1#2#3{Nucl. Phys. {\bf B#1} (#2) #3}
\def\pl#1#2#3{Phys. Lett. {\bf #1B} (#2) #3}
\def\prl#1#2#3{Phys. Rev. Lett.{\bf #1} (#2) #3}

\def\ijmp#1#2#3{Int. Jour. Mod. Phys. {\bf A#1} (#2) #3}

\def\comments#1{}

\def\QM{\Bbb{M}}

\def\half{{1\over 2}}

\def\tr{{\rm tr\ }}

\def\bra#1{{\langle}#1|}
\def\ket#1{|#1\rangle}

\def\Dslash{\rlap{\hskip0.2em/}D}

\def\CP{{\cal P}}

\def\a{\alpha}

\def\II{\relax{I\kern-.07em I}}

\def\IZ{\relax\ifmmode\mathchoice
{\hbox{\cmss Z\kern-.4em Z}}{\hbox{\cmss Z\kern-.4em Z}}
{\lower.9pt\hbox{\cmsss Z\kern-.4em Z}}
{\lower1.2pt\hbox{\cmsss Z\kern-.4em Z}}\else{\cmss Z\kern-.4em
Z}\fi}
\def\IB{\relax{\rm I\kern-.18em B}}
\def\IC{\bf C}
\def\ID{\relax{\rm I\kern-.18em D}}
\def\IE{\relax{\rm I\kern-.18em E}}
\def\IF{\relax{\rm I\kern-.18em F}}
\def\IG{\relax\hbox{$\inbar\kern-.3em{\rm G}$}}
\def\IGa{\relax\hbox{${\rm I}\kern-.18em\Gamma$}}
\def\IH{\relax{\rm I\kern-.18em H}}
\def\II{\relax{\rm I\kern-.18em I}}
\def\IK{\relax{\rm I\kern-.18em K}}
\def\IP{\relax{\rm I\kern-.18em P}}
%\def\IX{\relax{\rm X\kern-.01em X}}
%this doesn't work

\useblackboard
\def\IZ{\relax\Bbb{Z}}
\fi

\font\cmss=cmss10 \font\cmsss=cmss10 at 7pt
\def\IR{\relax{\rm I\kern-.18em R}}

\def\BR{\IR}
\def\BZ{\IZ}
\def\BR{\IR}
\def\BC{\IC}
\def\BM{\QM}

\def\tilde{\widetilde}

%%%%%%%%%%%%%%%%%%%%%%%%%%%%%%%%%%%%%%%%%%%%%%%%%%%%%%%%%%%%%%%%%%%%%%%%%%%%
%                    F I G U R E S                                         %
%%%%%%%%%%%%%%%%%%%%%%%%%%%%%%%%%%%%%%%%%%%%%%%%%%%%%%%%%%%%%%%%%%%%%%%%%%%%
%\figsfalse

\iffigs
  \input epsf
\else
  \message{No figures will be included.  See TeX file for more
information.}
\fi

%%% \iffigs
%%% \midinsert
%%% $$\vbox{\centerline{\epsfxsize=4in\epsfbox{figa.eps}}
%%% \centerline{Figure 1. $E_2$ and related theories.}}$$
%%% \endinsert
%%% \fi

%%%%%%%%%%%%%%%%%%%%%%%%%%%%%%%%%%%%%%%%%%%%%%%%%%%%%%%%%%%%%%%%%%%%%%%%%%%%
%                    Definitions from LaTeX                                %
%%%%%%%%%%%%%%%%%%%%%%%%%%%%%%%%%%%%%%%%%%%%%%%%%%%%%%%%%%%%%%%%%%%%%%%%%%%%

%%%
%%% All those have problems with Font \rm
%%%

\def\coth#1{{{\rm coth}{#1}}}
\def\tanh#1{{{\rm tanh}{#1}}}
\def\cosh#1{{{\rm cosh}{#1}}}
\def\sinh#1{{{\rm sinh}{#1}}}

\def\lim{{lim}}

%%%%%%%%%%%%%%%%%%%%%%%%%%%%%%%%%%%%%%%%%%%%%%%%%%%%%%%%%%%%%%%%%%%%%%%%%%%%
%                    My definitions                                        %
%%%%%%%%%%%%%%%%%%%%%%%%%%%%%%%%%%%%%%%%%%%%%%%%%%%%%%%%%%%%%%%%%%%%%%%%%%%%
\input epsf

\def\SUSY#1{{{\cal N}= {#1}}}                   % N=? SUSY
                             % [
                             % ]

\def\wdg{{\wedge}}                              % wedge product

                              % Wilson lines

                              % inverse
                           % O(x)

\def\MR#1{{{\BR}^{#1}}}               % Real numbers
               % Complex numbers

%%% \def\MR#1{{{\bf R}^{#1}}}               % Real numbers
%%% \def\MC#1{{{\bf C}^{#1}}}               % Complex numbers
\def\MR#1{{{\BR}^{#1}}}               % Real numbers
               % Complex numbers
\def\MS#1{{{\bf S}^{#1}}}               % Circle, sphere,...
               % disk, ball,...
               % Torus
\def\CP#1{{{\bf P}^{#1}}}              % CP
               % Ruled surface F_n

             % Patch
                    % line-bundle
\def\px#1{{\partial_{#1}}}              % derivative

                 % Left large bracket
                % Right large bracket
              % SL(*,Z)

\def\Id{{\bf I}}                             % identity matrix

\def\com#1#2{{\lbrack {#1},{#2} \rbrack}}      % commutator
               % anti-commutator

\def\ev#1{{\langle {#1} \rangle}}           % expectation value
\def\evtr#1{{\langle \tr{{#1}} \rangle}}    % expectation value of trace

      % trace
\def\trp#1{{{\rm tr}\{ {#1} \} }}            % trace
            % Trace
            % trace in a rep
            % Trace in a rep

\def\rep#1{{{\bf {#1}}}}                      % representation
\def\ImX{{{\rm Im}\,X}}                  % Imaginary
\def\ReX{{{\rm Re}\,X}}                  % Imaginary

\def\widebar#1{{\overline{#1}}}                    % Wide bar
\def\sg#1{{ {\bf \sigma}^{{#1}} }}                 % Pauli matrix

\def\Ol#1{{ {\cal O}({#1}) }}                      % correction O()

                      % Hodge star
                         % sign
\def\hepth#1{{\it hep-th/{#1}}}

%%%%%%%%%%%%%%%%%%%%%%%%%%%%%%%%%%%%%%%%%%%%%%%%%%%%%%%%%%%%%%%%%%%%%%%%%%%%
%                    Greek                                                 %
%%%%%%%%%%%%%%%%%%%%%%%%%%%%%%%%%%%%%%%%%%%%%%%%%%%%%%%%%%%%%%%%%%%%%%%%%%%%
\def\u{{\mu}}
\def\v{{\nu}}

\def\da{{\dot{\a}}}

\def\bchi{{\bar{\chi}}}

\def\bret{{\bar{\eta}}}

\def\brho{{\bar{\rho}}}

%%% \def\Dsh{{D\!\!\!\slash}}     % D slash

%%%%%%%%%%%%%%%%%%%%%%%%%%%%%%%%%%%%%%%%%%%%%%%%%%%%%%%%%%%%%%%%%%%%%%%%%%%%
%     Special Purpose  Definitions                                         %
%%%%%%%%%%%%%%%%%%%%%%%%%%%%%%%%%%%%%%%%%%%%%%%%%%%%%%%%%%%%%%%%%%%%%%%%%%%%
                            % 2\times 2  J

\def\hX{{\widehat{X}}}
\def\wpr{{{\cal G}}}                                % Wave function propagator

%%%%%%%%%%%%%%%%%%%%%%%%%%%%%%%%%%%%%%%%%%%%%%%%%%%%%%%%%%%%%%%%%%%%%%%%%%%%
%                    TITLE PAGE                                            %
%%%%%%%%%%%%%%%%%%%%%%%%%%%%%%%%%%%%%%%%%%%%%%%%%%%%%%%%%%%%%%%%%%%%%%%%%%%%

%%% \draftmode

%
\Title{ \vbox{\baselineskip12pt\hbox{hep-th/9705188, PUPT-1680}}}
{\vbox{
\centerline{ Higher  Loop Effects in  M(atrix) Orbifolds.}}}
\centerline{Ori J. Ganor, Rajesh  Gopakumar and Sanjaye Ramgoolam}
\smallskip
\smallskip
\centerline{Department of Physics, Jadwin Hall}
\centerline{Princeton University}
\centerline{Princeton, NJ 08544, USA}
\centerline{\tt origa,rgk,ramgoola@puhep1.princeton.edu}
%%%
\bigskip
\bigskip
\noindent

Scattering of zero branes off the fixed point 
in $R^8/Z_2$, as described by a super-quantum mechanics with 
eight supercharges, displays some novel effects 
relevant to Matrix theory in non-compact backgrounds.
The leading long distance behaviour of  the moduli space metric   
receives no correction at one loop in Matrix theory,
but does receive  a correction at   two loops.
There are no contributions  at  higher  loops. 
We explicitly calculate the two-loop term, finding a non-zero result. 
We find a discrepancy with M(atrix)-theory.
Although the result has the right dependence on $v$ and $b$
for the scattering of zero branes off the fixed point
the factors of $N$ do not match.
We also discuss scattering in the orbifolds,
$R^5/Z_2$ and $R^9/Z_2$ where we find
the predicted fractional charges. 
 
\Date{May, 1997}

%%%%%%%%%%%%%%%%%%%%%%%%%%%%%%%%%%%%%%%%%%%%%%%%%%%%%%%%%%%%%%%%%%%%
%  B I B L I O G R A P H Y                                         %
%%%%%%%%%%%%%%%%%%%%%%%%%%%%%%%%%%%%%%%%%%%%%%%%%%%%%%%%%%%%%%%%%%%%

\lref\BFSS{T. Banks, W. Fischler, S.H. Shenker and L. Susskind,
  {\it ``M Theory As A Matrix Model: A Conjecture,''} \hepth{9610043}}

\lref\Wati{W. Taylor,
  {\it ``D-brane field theory on compact spaces,''} \hepth{9611042}}

\lref\Sus{L. Susskind,
  {\it ``T Duality in M(atrix) Theory and S Duality in Field Theory,''}
  \hepth{9611164}}

\lref\BerDou{M. Berkooz and M.R. Douglas,
  {\it ``Five-branes in M(atrix) Theory,''} \hepth{9610236}}

\lref\DKPS{M.R. Douglas, D. Kabat, P. Pouliot and S. Shenker,
  {\it ``D-branes and Short Distances in String Theory,''}
     \hepth{9608024}}

\lref\BacPor{C.Bachas, M. Porrati,
   {\it ``Pair Creation of Open Strings in an Electric Field,''}
      \pl{296}{77}{92}, \hepth{9209032}}

\lref\Polpou{J.Polchinski, P.Pouliot,
   {\it ``Membrane Scattering with M-Momentum Transfer,''}
      \hepth{9704029}}

\lref\BL{V. Balasubramanian and F. Larsen,
  {\it ``Relativistic Brane Scattering,''} \hepth{9703039}}

\lref\DasMuk{K. Dasgupta and S. Mukhi,
  {``Orbifolds of M-Theory,''} 
      \np{465}{399}{96},\hepth{9512196}}

\lref\DasMukII{K. Dasgupta and S. Mukhi,
  {``A Note on Low-Dimensional String Compactifications,''}
     \hepth{9612188}}

\lref\SVW{S. Sethi, C. Vafa and E. Witten,
  {``Constraints On Low-Dimensional String Compactifications,''}
  \hepth{9606122}}

\lref\Russo{J.G. Russo,
  {\it ``BPS Bound States, Supermembranes and T-Duality
  in M-Theory,''} \hepth{9703118}}

\lref\Motl{L. Motl,
  {\it ``Proposals on Non-perturbative Superstring Interactions,''}
  \hepth{9701025}}

\lref\BM{T. Banks and L. Motl,
  {\it ``Heterotic Strings from Matrices,''} \hepth{9703218}}

\lref\Petr{Petr Horava,
  {\it ``Matrix Theory and Heterotic Strings on Tori''},
  \hepth{9705055}}

\lref\WitVAR{E. Witten,
  {``String Theory Dynamics in Various Dimensions,''}
  \np{443}{95}{85}, \hepth{9503124}}

\lref\WitORB{E. Witten,
  {``Five-Branes and M-Theory On an Orbifold,''}
  \np{463}{96}{383}, \hepth{9512219}}

\lref\DVV{R. Dijkgraaf, E. Verlinde, H. Verlinde,
  {\it ``Matrix String Theory,''} \hepth{9703030}}

\lref\KR{N. Kim and S.-J. Rey,
  {\it ``M(atrix) Theory on an Orbifold and Twisted Membrane,''}
  \hepth{9701139}}

\lref\BB{K. Becker and M. Becker,
   {\it ``A Two-loop test of M(atrix)-theory''}, \hepth{9705091}}

\lref\kapou{D. Kabat and P. Pouliot,
  {\it ``A Comment on Zero Brane Quantum Mechanics''},
  \hepth{9603127}, \prl{77}{96}{1004}}

\lref\DFS{U.H. Danielsson, G. Ferretti and B. Sundborg,
  {\it ``D-Particle Dynamics and Bound-States''},
  \hepth{9603081}, \ijmp{11}{96}{5463}}

\lref\DF{U.H. Danielsson and G. Ferretti,
  {\it ``The Heterotic Life OF The D-Particle,''}
  \hepth{9610082}}

\lref\DM{M. Douglas and G. Moore,
  {\it ``D-branes, Quivers And ALE Instantons''},
  \hepth{9603167}}

\lref\BB{K. Becker and M. Becker,
  {\it ``A Two-Loop Test of M(atrix) Theory''}, \hepth{9705091}}

\lref\EBS{T. Banks, N. Seiberg, and E. Silverstein,
  {\it ``Zero and One-dimensional Probes with N=8 Supersymmetry''},
  \hepth{9703052}}

% ===================================================================== %
% More Refs
% ===================================================================== %
\lref\Gop{R. Gopakumar,
  {\it ``BPS States In Matrix Strings''},
  \hepth{9704030}}

\lref\Sus{L. Susskind,
  {\it ``T Duality in M(atrix) Theory and S Duality in Field Theory,''}
  \hepth{9611164}}

\lref\SS{S. Sethi and L. Susskind,
  {\it ``Rotational Invariance in the M(atrix) Formulation
  of Type IIB Theory,''} \hepth{9702101}}

\lref\GRT{O.J. Ganor, S. Ramgoolam and W. Taylor IV,
  {\it ``Branes, Fluxes and Duality in M(atrix)-Theory,''} \hepth{9611202}}

\lref\KS{S. Kachru and E. Silverstein,
  {\it ``On Gauge Bosons in the Matrix Model Approach to M Theory,''}
  \hepth{9612162}}

\lref\BRS{M. Berkooz, M. Rozali and N. Seiberg,
  {\it ``Matrix Description of M-theory on $T^4$ and $T^5$,''}
  \hepth{9704089}}

\lref\Roz{M. Rozali,
  {\it ``Matrix Theory and U-duality in Seven Dimensions,''}
  \hepth{9702136}}

\lref\BS{T. Banks and N. Seiberg,
  {\it ``Strings from Matrices,''} \hepth{9702187}}

\lref\SetSte{S. Sethi and M. Stern,
  {\it ``D-Brane Bound States Redux,''} \hepth{9705046}}

\lref\AhBe{O. Aharony and M. Berkooz,
  {\it ``Membrane Dynamics in M(atrix) Theory,''} \hepth{9611215}}

\lref\GilSam{G. Lifschytz and S. Mathur,
  {\it ``Supersymmetry and Membrane Interactions in M(atrix) Theory,''}
  \hepth{9612087}}

\lref\gilad{G. Lifschytz,
  {\it ``Four-Brane and Six-Brane Interactions in M(atrix) Theory,''}
  \hepth{9612223}}

\lref\KRII{N. Kim and S.-J. Rey,
  {\it ``M(atrix) Theory on $T_5/Z_2$ Orbifold and Five-Brane''},
  \hepth{9705132}}

\lref\KRIII{N. Kim and S.-J. Rey, to appear.}

\lref\DGM{M.R. Douglas, B.R. Greene and D.R. Morrison,
  {\it ``Orbifold Resolution By D-Branes''}, \hepth{9704151}}

\lref\WitJC{E. Witten, explanation in the {\it Princeton
            Journal Club on the M(atrix)-conjecture}, March '97.}

\lref\WitPMF{E. Witten,
  {\it ``Phase Transitions In M-Theory And F-Theory,''}
  \hepth{9603150}}

\lref\WitCOM{ E. Witten,
  {\it ``Some Comments on String Dynamics,''}
  contributed to Strings '95, \hepth{9507121}}

\lref\VWtad{E. Witten and C. Vafa,
  {\it ``A One-Loop Test Of String Duality,''} \hepth{9505053}}

\lref\DOS{M. R. Douglas, H. Ooguri and S. H. Shenker,
  {\it ``Issues in (M)atrix Model Compactification,''}
  \hepth{9702203}}

\lref\Lowe{D.A. Lowe,
  {\it ``Heterotic Matrix String Theory,''} \hepth{9704041}}

\lref\DougI{M. R. Douglas,
  {\it ``Enhanced Gauge Symmetry in M(atrix) Theory,''}
  \hepth{9612126}}

\lref\DougII{M. R. Douglas,
  {\it ``D-Branes in Curved Space,''} \hepth{9703056}}

\lref\DistBC{David Berenstein, Richard Corrado, Jacques Distler,
  {\it ``On the Moduli Spaces of M(atrix)-Theory Compactifications''}
   \hepth{9704087} }

\lref\SJR{S.-J. Rey,
  {\it ``Heterotic M(atrix) Strings and Their Interactions,''}
  \hepth{9704158}}

\lref\BRII{M. Berkooz and M. Rozali,
  {\it ``String Dualities from Matrix Theory,''}
  \hepth{9705175}}

\lref\SusII{L. Susskind,
  {\it ``Another Conjecture about M(atrix) Theory,''} \hepth{9704080}}

\lref\FR{W. Fischler and A. Rajaraman,
  {\it ``M(atrix) String Theory on K3,''} \hepth{9704123}}

\lref\FHRS{W. Fischler, E. Halyo, A. Rajaraman, and L. Susskind,
  {\it ``The Incredible Shrinking Torus,''} \hepth{9703102}}

\lref\BC{D. Berenstein and R. Corrado,
  {\it ``M(atrix)-Theory in Various Dimensions,''} \hepth{9702108}}

\lref\GR{Z. Guralnik and S. Ramgoolam,
  {\it ``Torons and D-Brane Bound States,''} \hepth{9702099}}

\lref\BCD{D. Berenstein, R. Corrado, and J. Distler,
  {\it ``On the Moduli Spaces of M(atrix)-Theory Compactifications,''}
  \hepth{9704087}}

\lref\FS{A. Fayyazuddin and D.J. Smith,
  {\it ``A note on $T^5/Z_2$ compactification 
  of the M-theory matrix model,''} \hepth{9703208}}

\lref\BSS{T. Banks, N. Seiberg and S. Shenker,
  {\it ``Branes from Matrices,''} \hepth{9612157}}

% ===================================================================== %
% Introduction
% ===================================================================== %
\newsec{Introduction}

The boldness of the proposal for an exact formulation for 11D M-theory 
\BFSS\ has provoked a rather intensive comparison of this theory
with known limits of M-theory. Many studies have been undertaken of simple
compactifications and properties of BPS states therein. 
\refs{\BFSS,\Wati,\Sus,\GRT,\DF,\KS,\KR,\GR,\SS,
\BC,\Roz,\FHRS,\BSS,\Gop,\BCD,\BRS,\FR}
The type-IIA string theory limit \refs{\BS,\Motl,\DVV}
together with interactions \DVV\ has been argued to arise naturally.
The perturbative heterotic theory and its compactifications
have also been recently studied \refs{\BM,\Lowe,\SJR,\Petr}.

Another line that has been pursued is the traditional one of scattering
objects (mostly BPS) off each other.  
It was one of the interesting features of the original proposal that the 
``tree-level'' ${{v^4}\over {b^7}}$ effective potential between
gravitons in M-theory was reproduced by a 1-loop effect in the
0+1D SYM \BFSS. The leading small 
velocity and long distance behaviour in the
scattering of many brane configurations are also reproduced exactly 
by 1-loop calculations \refs{\AhBe,\GilSam,\gilad}. 
An interesting calculation for a process with non-zero longitudinal momentum 
transfer has also been carried out in \Polpou and yields the correct result. 

One motivation for the present paper was to study
the loop expansion of M(atrix)-theory.
It is natural to look for effects in M(atrix)-theory that can only be
seen at higher loops in the Quantum Mechanics. This could be particularly
educative in cases with less than maximal supersymmetry. 
As we will see, such a  situation in which we can see a {\it qualitative}
effect only by going to two loops,
arises in the scattering off the fixed point of $\MR{8}/\BZ_2$.
In M-theory language, what we are probing is a quantum effect 
-- the effective fractional
membrane charge at the fixed point which is related to the 
tadpole discovered in \VWtad.
In principle, the details of the bound-state in the large-$N$ limit
could contribute as well. Its details are mostly unknown and in fact, 
it is only recently, that the existence of an $SU(2)$ bound state has been
proven rigorously \SetSte. We will show that
the details of the bound state wave-function will contribute only
at subleading order.

A parallel motive for this work was to understand something about 
the crucial issue of compactifications on nontrivial backgrounds.
The prescription given in \BFSS\ for toroidal compactification
was to use ``large" gauge transformations of the model.
We recall that when $N=\infty$ states have to be invariant
under ``small" gauge transformations, i.e. $U\in U(\infty)$ such
that $\| U-\Id \| < \infty$. By using appropriate ``large"
gauge transformations (i.e. $U\in U(\infty)$ such that
$\| U-\Id \| = \infty$) one can map toroidal compactifications
to SYM theories in various dimensions \refs{\BFSS,\Wati,\Sus,\GRT}.
Nontoroidal compactifications have been studied in
\refs{\DougI,\DOS}.

There seem to be several problems with such compactifications.
In general, compactifications to less that eight dimensions seem
to contain a gauge theory in more than 3+1D and those are not
renormalizable (see \BS\ for a discussion on this subject).
In compactifications to lower dimensions it seems necessary to
use the newly discovered \WitCOM\ chiral theories in 5+1D 
\refs{\Roz,\BRS,\BRII}
but these do not seem to appear naturally from matrices.

Therefore, we do not have at the moment a prescription
for how to formulate M(atrix) theory on a general background.
The orbifold limits of compactifications have been formulated
according to the prescription of \BFSS\ by restricting to
the part of the parameter space that is invariant under
``large'' gauge transformations. The authors of \DOS\ have
pointed out several problems with this model
in deforming away from the orbifold point and have shown
that compactifications on K3
cannot be reproduced by a finite number of ``off-diagonal'' degrees
of freedom.

In order to ``isolate'' the problems, it seems worthwhile
to study the non-compact orbifolds $\MR{d}/\Gamma$.
These would be given by a 0+1D theory with less than 16 supercharges.
The examples that we study have (excepting one) 8 supercharges.
General theories with 8 supersymmetries in 0+1D have
been recently studied in \EBS\ and indeed, supersymmetry allows
for higher loop corrections to the metric.\foot{We are grateful
to N. Seiberg for bringing this to our attention.}

Our paper is organized as follows.
Section (2) is a review of effects near orbifolds in M-theory.
We review the M-theory quantum effects near
$\MR{4}/\BZ_2$, $\MR{5}/\BZ_2$, $\MR{8}/\BZ_2$ and $\MR{9}/\BZ_2$.
We calculate the scattering off $\MR{5}/\BZ_2$ and $\MR{9}/\BZ_2$
in M(atrix)-theory and briefly discuss the scattering on orbifolds
which accommodate non-trivial low-energy theories: $\MR{4}/\BZ_2$
(free in the IR) and $\MR{6}/\BZ_3$ (interacting). 
(Such theories have been extensively discussed in \DGM\ from a different
perspective.)
In section (3) we explain why we expect the two-loop diagrams
to be the leading contribution in the $\MR{8}/\BZ_2$ case. 
We discuss the general behaviour of multi-loop diagrams. 
Our discussion implies, for example, that 
 in the $SU(N)$ quantum mechanics at finite $N$, higher loops
do not affect the ${v^4 \over {r^7} }$ term in the effective potential. 
In section (4) we list the technical details of the two-loop calculation
near $\MR{8}/\BZ_2$.
In section (5) we look at the wave-function contribution and
argue that the leading term
is ``universal'' in the sense that it can be determined
without a detailed knowledge of the bound-state wave-function.
In section (6) we present the final result and in section (7)
we discuss the large $N$ behaviour.

%=======================================================================%
% Note added
%=======================================================================%
\bigbreak\bigskip\bigskip
\centerline{\bf Note added}\nobreak
As this work neared completion, a paper which uses similar two-loop
techniques in a related context  appeared \BB.

We have also learnt about another work \refs{\KRII,\KRIII}
which has calculated scattering off $\MR{5}/\BZ_2$ and $\MR{9}/\BZ_2$.
We are grateful to the author for discussions.

% ===================================================================== %
% Section (2): Quantum effects at orbifolds
% ===================================================================== %
\newsec{Quantum effects at orbifolds}

In M-theory there are quantum effects at various orbifolds. For example,
The $\MR{5}/\BZ_2'$ orbifold \refs{\DasMuk,\WitORB}
(the $\BZ_2'$ includes the $A_3\rightarrow -A_3$
transformation) is required to have $-\half$ a 
5-brane charge at the fixed point \WitORB. This means that at large distances
the low-energy gravitational fields are those that would be induced
around an object with the mass and charge of $\half$ an anti 5-brane.
Similarly, at the $\MR{8}/\BZ_2$ fixed point there is, {\it effectively},
$-{1\over {16}}$ of a membrane charge.
There are other orbifolds in which the low-energy description 
requires a non-trivial theory at the fixed points.
For $\MR{4}/\BZ_2$ this is the $SU(2)$ gauge theory at an $A_1$
singularity \WitVAR, which is non-trivial but free in the IR.
$\MR{9}/\BZ_2'$ has a 1+1D CFT \DasMukII.
After compactification on $\MS{1}$
we find as a result $-{1\over {32}}$ of an anomalous
momentum (0-brane charge)\DasMukII.
There are more complicated examples, like $\MR{6}/\BZ_3$ where
the singularity involves a collapse of a $\CP{2}$ and a non-trivial
5D IR fixed point \WitPMF.

These effects influence the scattering of gravitons off the fixed
points and we are going to discuss the calculation both in the
low-energy supergravity and in the quantum mechanics.
The approximation that is made here is that of small momenta for the
graviton in the orbifolded dimensions.

In this section we calculate the scattering off
$\MR{5}/\BZ_2'$ and $\MR{9}/\BZ_2$ which involve a 1-loop effect 
in the gauge quantum mechanics.
In the next sections we will calculate the 
scattering off $\MR{8}/\BZ_2$ which exhibits interesting
2-loop effects.

% --------------------------------------------------------------------- %
% R^5/Z_2
% --------------------------------------------------------------------- %
\subsec{Scattering off $\MR{5}/\BZ_2'$}

The M(atrix)-model for $\MR{5}/\BZ_2$ has been given in \refs{\KR,\FS}. 

The 0+1D Lagrangian has an $Sp(N)$ gauge group.
There is one gauge multiplet which contains the gauge field $A_0$, its
 5 bosonic superpartners $X_\u$ ($\u=1\dots 5$)
and 8 fermionic ``gluino'' superpartners $\theta_{\a i}$
with $\a=1\dots 4$ a spinor index of the transverse $SO(5)$ and $i=1,2$.
The fields $(A_0, X_\u, \theta_{\a i})$ are in the adjoint $\rep{N(2N+1)}$
of $Sp(N)$.
The other multiplet contains 4 complex bosonic and 4 complex fermionic
fields $(\Phi_a, \bar{\Phi}^a, \psi_\a)$ ($a=6\dots 9$)
where $\psi_\a$ ($\a=1\dots 4$)  is a spinor of $SO(5)_{(1\dots 5)}$.
They are in the antisymmetric
$\rep{N(2N-1)}$ of $Sp(N)$ and correspond to the $6\dots 9$
space-time coordinates.
 
Using similar techniques as in 
\DKPS\ \GilSam we find the phase shift for scattering one D0-brane
off the fixed point at impact parameter $b$, velocity $v$. The one loop
determinantal contribution is:
\eqn\fdkps{\eqalign{
&
{\det}^{-2} (-\px{\tau}^2 + \gamma^2\tau^2 + b^2)
{\det}^{-1} (-\px{\tau}^2 + \gamma^2\tau^2 + b^2 + 2 \gamma) \cr
&
{\det}^{-1} (-\px{\tau}^2 + \gamma^2\tau^2 + b^2 - 2\gamma)
{\det}^4\pmatrix{ \px{\tau} & \gamma\tau - ib \cr
                 \gamma\tau + ib & \px{\tau} \cr }. \cr
}}
where $\tau = it$ and $\gamma = -i v$.
Note that we have 4 transverse bosons instead of 8 for $\SUSY{4}$
and 8 fermions instead of 16.

In the limit $v/b^2 \rightarrow 0$ one finds on taking the logarithm,
that the phase shift is
$$
\delta{(b,v)} = -{v\over {2b^2}}
$$
Remembering that the fixed point charge in the covering space 
(where we have been working) is twice 
that of the orbifold, this phase shift is in
agreement \DKPS\ with half a 5-brane at the orientifold point.
Thus, at 1-loop level M(atrix)-theory gives the same result,
since in the type-IIA diagrams, the contribution of full SUSY
multiplets cancels \BacPor\BFSS.

% --------------------------------------------------------------------- %
% R^9/Z_2
% --------------------------------------------------------------------- %
\subsec{Scattering off $\MR{9}/\BZ_2$}

The next example that we will discuss is M-theory on 
$\MR{1,1}\times(\MR{9}/\BZ_2')$.
This is a more complicated effect because of the nontrivial CFT
that lives on the fixed point.
There is only one homogeneous space dimension and
the anomalous charge that is associated with
the orbifold is a fractional
unit of momentum which appears whenever we compactify the homogeneous
direction on a large $\MS{1}$ (and disappears in the infinite radius
limit).
This is the signature of a nontrivial 1+1D CFT.
The fraction of momentum is determined to be $-{1\over {32}}$ \DasMukII.

The M(atrix) model for M-theory on $\MR{9}/\BZ_2'$ can be
constructed as in \KR. It is the dimensional reduction to 0+1D
of $SO(2N)$ Yang-Mills in 10D. This model
has the correct moduli space of $(\MR{9}/\BZ_2)^N/S_N$ but
at first sight this twice as many supersymmetries. However, the
nonlinearly realized supersymmetries of \BFSS, i.e.

$$
\delta\theta = \epsilon\Id,
$$
are no longer there because $SO(2N)$ doesn't have a $U(1)$ center
like $U(N)$.

We will recover
the $-{1\over {32}}$ quantum charge only in the type-IIA limit.
We scatter one 
D0-brane off the orientifold point, as before at impact parameter $b$
and velocity $v$.
The model is $SO(2)$ SYM and so there are no interactions.
On the other hand the D0-brane interacts with its image
and gives a force (gradient of the potential) of:
$$
F_1(b)=7{{v^4}\over {(2b)^8}}
$$
since $2b$ is the distance to the image brane.
This has to be canceled by an effective D0-brane charge $Q$ ($2Q$ in
the covering space), 
at the fixed point which contributes
$$
F_2(b)=2 Q\times 7 {{(\half v)^4}\over {b^8}}
$$
note that the relative velocity is $\half v$ since the orientifold is
fixed. Requiring the two contributions to cancel gives us precisely
$Q=-{1\over {32}}$.

In the M-theory limit, when the size of the 11th dimension
is much larger that the impact parameter $b$, a graviton
doesn't feel a force from the fixed point, only from its image.
This agrees with M(atrix)-theory since in the large $N$ limit
the potential between the D0-brane and its image behaves like $N^2$
whereas the potential between the D0-brane and fixed point behaves
only like $N$.

% --------------------------------------------------------------------- %
% R^8/Z_2
% --------------------------------------------------------------------- %
\subsec{Scattering off a membrane in 11D supergravity}

Our main focus in this paper is the scattering off 
$\MR{8}/\BZ_2$ which we will describe in detail in later sections.
In this subsection we will simply calculate the expected
classical supergravity
result for the scattering phase shift off  a membrane.
We do this to leading order in $v$ and ${1\over b}$.
We write down the
classical solution  of a membrane in 10+1D, 
see for example \Russo:
\eqn\memb{\eqalign{
ds^2 &= H^{-2/3} (-dt^2 + dy_1^2 + dy_{11}^2)
         + H^{1/3} dx_i dx^i,\cr
C_3 &= H^{-1} dt\wdg dy_1\wdg dy_{11},\cr
H &= 1 + {Q_2\over {r^6}}\cr
}}
where $Q_2= 8N_2 (2\pi)^2 (l_p)^6$ and $N_2$ is an integer. 
Using
$$
|\vec{p}| {{dp^i}\over {dt}}
= {\Gamma^i}_{00} |\vec{p}|^2 + {\Gamma^i}_{jj} (p^j)^2
$$
the geodesic equation for a graviton in this background is:
$$
|\vec{p}| {{dp_3}\over {dt}}
 = \Gamma_{300} |\vec{p}|^2 
 + \Gamma_{333} (p^3)^3 + \Gamma_{3,11,11} (p^{11})^2
$$
with
$$
p_3 = {{N_0 v}\over {R_{11}}},\qquad
p_{11} = {{N_0}\over {R_{11}}},
\Longrightarrow
|\vec{p}| = {{N_0}\over {R_{11}}} (1 + v^2)^{1/2}.
$$
This gives
$$
|\vec{p}| {{dp^i}\over {dt}}
= - {{Q_2 x_3}\over {r^8}} \left\lbrack
2|\vec{p}|^2 + p_3^2 - 2 |p_{11}|^2
\right\rbrack = -{{3 Q_2 x_3}\over {r^8}} (p_3)^2.
$$
So
$$
{{\partial V}\over {\partial x_3}} =
{{3 Q_2 x_3 v^2 N_0} \over {r^8 R_{11}}}
\Longrightarrow
V(r) = - {{Q_2 N_0 v^2}\over {2 r^6 R_{11}}}
$$
which corresponds to a  phase shift of 
$$ 
\delta = {{3\pi v Q_2 N_0}\over {16 R_{11} b^5}}=
    {  3 (2\pi)^3 v N_2N_0 l_p^6 \over {2b^5 R_{11} }}  
$$

To compare with the phase shift to be calculated in the gauge theory, 
we need to rescale 
$$
b \rightarrow b l_p (2\pi)^{1/3},\qquad
 v \rightarrow v {R_{11}\over l_p}
 (2\pi)^{-1/3}.
$$
(This is the rescaling that takes one from the 0-brane Lagrangian
in 11 dimensional units to that used in Section 4.)
This gives
\eqn\phsf{
\delta = {{3\pi v N_2 N_0}\over {2 b^5}}
}

% --------------------------------------------------------------------- %
% Nontrivial fixed points
% --------------------------------------------------------------------- %
\subsec{Comments on scattering off nontrivial fixed points}

The phenomena that we have studied so far are just effective
charges (except $\MR{9}/\BZ_2'$).
There are more complicated orbifolds where extra degrees of freedom
appear at the fixed points. 
This is the case of $\MR{4}/\BZ_2$ \WitVAR\ where the 
IR theory at the orbifold has 3 vector multiplets for the $W^0$
and $W^\pm$ bosons. An even more interesting example is $\MR{6}/\BZ_3$
where the low-energy theory is a non-trivial interacting fixed point.

Whenever we have such extra degrees of freedom, there are
nontrivial correlators of the energy-momentum tensor
$$
\ev{T_{\u_1 \v_1}(x_1) T_{\u_2 \v_2}(x_2) \cdots T_{\u_n \v_n}(x_n)}.
$$
where $x_i$ are localized at the fixed point.
The lowest order effect will be an addition of
\eqn\addof{
\int d^l q K_{ijkl}(q) h_{ij}(-q,0) h_{kl}(q,0),
}
to the action, where 
\eqn\gandk{\eqalign{
g_{\u\v} &= \eta_{\u\v} +
\int e^{i q_\Vert \cdot x_\Vert + i q_\perp \cdot x_\perp}
h_{\u\v}(q_\Vert, q_\perp),\cr
K_{ijkl}(q_\Vert) &= \ev{T_{ij}(q_\Vert) T_{kl}(-q_\Vert)},\cr
}}
Here $p_\Vert$ is the momentum in the homogeneous directions
(6+1 directions for $\MR{4}/\BZ_2$ and 4+1 for $\MR{6}/\BZ_3$).
To lowest order the amplitude doesn't depend on the perpendicular
momentum $p_\perp$ (in the 4 directions for $\MR{4}/\BZ_2$
and 6 directions for  $\MR{6}/\BZ_3$) because $T_{\u\v}(x)$
contains a $\delta$-function $\delta(x_\perp)$.
By conformal invariance we expect
\eqn\kcal{\eqalign{
K_{ijkl}(q)
&= {c\over {q^4}}\{(\delta_{ik}\delta_{jl}+\delta_{il}\delta_{jk}
-{2\over D}\delta_{ij}\delta_{kl}) q^4 \cr
&-q^2 (q_i q_k \delta_{jl} + q_i q_l \delta_{jk} + q_j q_l \delta_{ik}
 + q_j q_k \delta_{jl}) \cr
&+{2\over D}q^2 (q_i q_j \delta_{kl} + q_l q_k \delta_{ij})
+ (2-{2\over D}) q_i q_j q_k q_l \} \cr
}}
where $c$ is a generalization of the central charge.

Nevertheless, \addof\ cannot be used to calculate scattering
of a graviton off the fixed point. This is because \addof\
contains $\delta(x_\Vert)$ and to solve for a graviton plain
wave one needs an extra information about the ``profile''
of the $\delta$-function (in analogy to scattering off
a $\delta$-function potential in quantum-mechanics).

% ===================================================================== %
% Section (3): Behaviour of multi-loop diagrams
% ===================================================================== %
\newsec{Behaviour of multi-loop diagrams}

The cases discussed in the previous section were all tree-level
effects in supergravity. In contrast, the scattering off $\MR{8}/\BZ_2$,
or the effective membrane charge, can be deduced from a tadpole in 
a 1-loop diagram in type-IIA string theory \VWtad.
Closed string 
tree-level in string theory is related to 1-loop in M(atrix)-theory,
and  1-loop in string theory should be related to 2-loop in M(atrix)-theory.

In the next sections we will explicitly
perform the calculation.
In this section we wish to analyze, on general grounds, the behaviour
of multi-loop diagrams.

M theory in eleven dimensions has no free parameters, once we have 
set the Planck scale to $1$. Perturbation theory is possible 
because we are looking at low energy, long distance scattering. 
 In the DKPS formalism \DKPS\ this corresponds to large impact
parameter $b$ and low velocity $v$.
Similarly, the BFSS Hamiltonian, after an appropriate rescaling
of time, has no coupling constants. These facts indicate 
that the loop expansion we are doing is really an expansion 
depending on the physical parameters of the problem.
To be precise, we  show  that $L$-loop diagrams behave as
$$
{1\over {b^{3(L-1)}}} f_L({v\over {b^2}}).$$
The same argument  will apply to the non-compact orbifolds 
we are considering, since there are no extra compactification
parameters.

We write down the zero brane quantum mechanics 
Lagrangian : 
\eqn\zerob{\eqalign{ 
 S &= { (\alpha^{\prime})^{2} \over {g \sqrt{ \alpha^{\prime}}}}   
      \int  d\tau \left\lbrack\half (D_{\tau} \Phi_I  )^2 + 
       {1\over 4}[\Phi_{I}, \Phi_{J}]^2 
 + i  \Psi^T D_{\tau} \Psi  - \Psi^T\Gamma^{I} [\Phi_I, \Psi]
   \right\rbrack\cr 
 S &= {1\over {R^3}} 
    \int d\tau  \left\lbrack {1\over 2} (D_{\tau} \Phi_I  )^2 + 
         {1\over 4} [\Phi_{I}, \Phi_{J}]^2
 +       i \Psi^T D_{\tau} \Psi
      - \Psi^T\Gamma^{I} [\Phi_I, \Psi] \right\rbrack\cr }}
In going to the second line we used 
\eqn\alph{
 \alpha^{\prime} = { l_p^3\over {R}},\qquad
  g\sqrt{\alpha^{\prime}} = R 
}
where $R $ is the eleven dimensional radius and we set 
$l_p=1$.  
Performing perturbation theory in this language 
weights a diagram with $L$ loops by a factor $R^{3(L-1)}$, 
since $R^3$ plays the role of Planck's constant.
There is a rescaling of field variables and time which gets 
rid of the $R$ dependence completely \refs{\kapou,\DFS}. Indeed, define
\eqn\resc{\eqalign{ 
 & X = R^{-1} \Phi \cr 
 & \tau = { t \over R } \cr 
 & \Theta = R^{-3/2}\Psi \cr }}
In terms of these new variables the Lagrangian is 
\eqn\zrmath{
S = \int dt  \left\lbrack {1 \over 2 } (D_{t}  X_I  )^2 + 
     {1\over 4} \com{X_{I}}{X_{J}}^2  + 
      i \Theta^T D_{t} \Theta
      - \Theta^T\Gamma^{I} \com{X_I}{\Theta}
     \right\rbrack }
Now we will expand around the backgrounds 
\eqn\back{\eqalign{ 
& x_2 = vt  \cr
&  x_3 = b.   \cr  }}
 The above rescalings 
 allow us to relate background variables
\eqn\othb{\eqalign{ 
& \phi_2 = \tilde v \tau \cr  
& \phi_b = \tilde b \cr } }
to those in the $x$ variables. 
\eqn\rescb{\eqalign{ 
 &  \tilde b = R v \cr 
 & \tilde v = R^2 v \cr }} 
Let $\delta^{(L)} (b,v) $ be the phase shift computed with the 
Lagrangian \zrmath\ with backgrounds \back, and 
$(\delta^{\prime})^{(L)}  (b,v) $ 
be the phase shift computed with 
\zerob\ and the backgrounds \othb. 
They are related as 
\eqn\relexp{   \delta^{(L)} (b,v) = 
  R^{3(L-1)} (\delta^{\prime})^{(L)}  (\tilde b, \tilde v)
} 
This is solved by a phase shift of the 
form 
\eqn\solcon{ \delta^{(L)} (b,v) = b^{-3 (L-1)} f_L ( {v \over{ b^2}} ) } 
So the one loop answers are purely functions 
of ${v\over {b^2}}$.  
 
  This  argument  shows that higher loops cannot give 
   a correction to the ${ v^3 \over {b^6}}$  term
   of the phase shift in the scattering of two zero  branes. 
   It allows terms proportional to $v^3$ but necessarily 
    suppressed by higher powers of $b$ than $b^6$. 
   Non perturbative effects will give factors 
   $e^{-b^3}$, hence will not affect the coefficient 
   of the ${ v^3 \over {b^6}}$.
Similarly, it is clear that we cannot get corrections to the 
${v\over b^5}$ term by higher than two-loop calculations.

% --------------------------------------------------------------------- %
% Dependence on N
% --------------------------------------------------------------------- %
%%% \subsec{ Dependences on $N$. }

The same argument can be generalized to the case of 
the $U(N)$ theory in the original BFSS proposal ($U(N)\times U(N)$ for
the $\MR{8}/\BZ_2$  case).
Large N counting shows that
the $L$-loop diagrams behave, at leading order in large $N$,  as 
\eqn\NFL{
 {{N^{L+1}}\over {b^{3(L-1)}}} f_L ( {v\over {b^2}} ).
}
This is the large $N$ generalization of \solcon.
There are also diagrams with subleading dependence ($\sim N^{L+1-\rho}$)
on $N$.
This is discussed in more detail in section (7).

%%%    The previous argument also applies to the exact evaluation of the 
%%%    path integral around the background 
%%%    $X_2 = vt \sigma_3 \otimes \Id_{N \times N}$, 
%%%    $X_3 =  b \sigma_3 \otimes \Id_{N\times N} $, 
%%%     where these are $N\times N$  matrices. 
%%% 
%%% Note that the explicit calculations in this 
%%% paper are done for $U(1)\times U(1)$ but the 
%%% same argument above can also be applied 
%%% if we do the calculation of the path integral
%%% in the $U(N)\times U(N)$ theory of $N$ zero branes around 
%%% the background: 
%%% \eqn\nonab{\eqalign{ 
%%%   & X_2 = b \otimes \Id_{N\times N}\cr 
%%%   & X_3 = vt\otimes \Id_{N\times N}  \cr }} 
%%% where $X_2$ and $X_3$ are  $N\times N$ matrices proportional 
%%% to the identity. It also applies to the analogous calculation
%%% in the case of $N$ zero branes. 

% ===================================================================== %
% Section (4): Technical details
% ===================================================================== %
\newsec{Technical details}

In this section we will describe in detail the two-loop calculation.

% --------------------------------------------------------------------- %
% The Lagrangian
% --------------------------------------------------------------------- %
\subsec{The M(atrix)-Model for $\MR{8}/\BZ_2$}

Following \refs{\DM,\KR} we obtain the M(atrix)-Model for 
$\MR{8}/\BZ_2$ by starting with a $U(2N)$ SYM \zrmath, picking
the $U(2N)$ matrix:
$$
U = \pmatrix{\Id_{N\times N} & 0 \cr 0 & -\Id_{N\times N} \cr}
$$
and leaving only those modes which satisfy
\eqn\leav{\eqalign{
\widetilde{A}_0 &= U^{-1} \widetilde{A}_0 U,\cr
\widetilde{X}_1 &= U^{-1} \widetilde{X}_1 U,\cr
\widetilde{X}_I &= -U^{-1} \widetilde{X}_I U,\qquad I=2\dots 9,\cr
\widetilde{\Psi} &= \Gamma^{2}\cdots \Gamma^{9} U^{-1}\widetilde{\Psi} U.\cr
}}
We obtain a $U(N)\times U(N)$ gauge
group, one bosonic field $X_1$ in the adjoint, and 8 bosonic fields
$X_2\dots X_9$ in the $(\rep{N},\rep{\widebar{N}})$.
In terms of these the original $U(2N)$ fields are 
\eqn\interms{\eqalign{
\widetilde{A}_0 &= \pmatrix{A_0 & 0 \cr 0 & A_0' \cr},\cr
\widetilde{X}_1 &= \pmatrix{X_1 & 0 \cr 0 & X_1' \cr},\cr
\widetilde{X}_a &= \pmatrix{0 & X_a \cr X_a^\dagger & 0 \cr},
\qquad a=2\dots 9,\cr
}}
We will denote the two $U(N)$ gauge fields by $A_0,A'_0$
and the two bosonic fields corresponding to the two $U(N)$-s
by $X_1,X_1'$.
The quotienting prescription \leav\ leaves
8 fermions in the $\rep{8}_s$ of $SO(8)$ and
in the adjoint of $U(N)\times U(N)$ and additional
fermions $\psi^\da$
in the $\rep{8}_c$ of $SO(8)$ and in the $(\rep{N},\rep{\bar{N}})$
of $U(N)\times U(N)$. 

The explicit calculations will be  carried out
for $U(1)\times U(1)$.
The bosonic background will be taken to be:
\eqn\nonab{\eqalign{ 
  & X_2 = vt,\cr
  & X_3 = b.\cr 
}} 
We decompose 
$$
\psi^\da = \chi^\da + i\rho^\da,\qquad
$$
where $\chi$ and $\rho$ are real.
We group them as follows:
\eqn\etxi{
\xi^i = \pmatrix{ {\psi^i}' + \psi^i \cr 2\bchi^i \cr },\qquad
\eta^i = \pmatrix{ {\psi^i}' - \psi^i \cr 2\brho^i \cr }.
}
We also define the bosonic fields $\Phi_{\pm}$  according to
$$
\ImX_2 = {{\Phi_{+} - \Phi_{-}} \over {\sqrt{2}}},\qquad
A_0 - A_0' = \sqrt{2}(\Phi_{+} + \Phi_{-})
$$
The Lagrangian is now a sum of bosonic, fermionic and ghost
terms
$$
L = L_b + L_f + L_g
$$
The bosonic terms are
$$
L_b = L_b^{(2)} + L_b^{(3)} + L_b^{(4)}
$$
with the quadratic terms:
%%% \eqn\Lbstwo{\eqalign{
%%% L_b^{(2)} &=
%%%  (\px{\tau}(\ImX_a))^2  + 4 (b^2 +\gamma^2\tau^2) (\ImX_a)^2 
%%% +(\px{\tau}(\ImX_3))^2  + 4 (b^2 +\gamma^2\tau^2) (\ImX_3)^2
%%% \cr
%%% &+(\px{\tau}(\Phi_{+}))^2  + 4 (b^2 +\gamma^2\tau^2 + \gamma) (\Phi_{+})^2 
%%% +(\px{\tau}(\Phi_{-}))^2  + 4 (b^2 +\gamma^2\tau^2 - \gamma) (\Phi_{-})^2 
%%% \cr
%%% &+{1\over 4}\lbrack\px{\tau}(X_1 - X_1')\rbrack^2  
%%% + (b^2 +\gamma^2\tau^2) (X_1 - X_1')^2 
%%% + {1\over 4}\lbrack\px{\tau}(A_0 + A_0')\rbrack^2 
%%% \cr
%%% }}
\eqn\Lbstwo{\eqalign{
L_b^{(2)} &=
(\px{t}\ReX_a)^2 + (\px{t}\ReX_2)^2 + (\px{t}\ReX_3)^2
\cr
&+{1\over 4}\lbrack\px{t}(X_1 + X_1')\rbrack^2
+ {1\over 4}\lbrack\px{t}(A_0 + A_0')\rbrack^2 
\cr
&+(\px{t}\ImX_a)^2  - 4 (b^2 + v^2 t^2) (\ImX_a)^2 
+(\px{t}\ImX_3)^2  - 4 (b^2 + v^2 t^2) (\ImX_3)^2
\cr
&+(\px{t}\Phi_{+})^2  - 4 (b^2 + v^2 t^2 + v) \Phi_{+}^2 
+(\px{t}\Phi_{-})^2  - 4 (b^2 + v^2 t^2 - v) \Phi_{-}^2 
\cr
&+{1\over 4}\lbrack\px{t}(X_1 - X_1')\rbrack^2  
- (b^2 + v^2 t^2) (X_1 - X_1')^2 
\cr
}}
The cubic and quartic terms $L_b^{(3)}, L_b^{(4)}$ appear
in the appendix.
The fermionic quadratic terms are
\eqn\lfquad{
L_f =
i\xi_i^\dagger \px{t}\xi^i + i\eta_i^\dagger \px{t}\eta^i
+2 v t \eta_i^\dagger\sg{2}\eta^i + 2 b \eta_i^\dagger\sg{1}\eta^i.
}
The cubic terms with fermions also appear in the appendix.

% --------------------------------------------------------------------- %
% One-loop
% --------------------------------------------------------------------- %
\subsec{One-loop}

The one-loop contribution to the phase-shift is given by:
$$
\delta(b,v) =
\int {ds\over s} \times {1\over {2\sinh{2 s v}}} e^{-4 b^2 s}
\left\lbrack
6 + 2 \cosh{4 s v} - 8 \cosh{ 2 s v}
\right\rbrack = {{v^3}\over {16 b^6}} + \Ol{{{v^5}\over {b^{10}}}},
$$
similarly to \DKPS.

% --------------------------------------------------------------------- %
% Diagrams and propagators
% --------------------------------------------------------------------- %
\subsec{Diagrams and propagators}

The propagators of the bosonic and fermionic
fields can be read off from the kinetic operators
in \Lbstwo\ and \lfquad\
(analytically continued to a Euclidean metric):
$$
\Dslash \equiv
i\sg{3}{d\over{dt}} - 2 v t\sg{1} + 2 b \sg{2},\qquad
H \equiv -{{d^2}\over {dt^2}} + 4 v^2 t^2 + 4 b^2.
$$
we find
$$
\Dslash^2 = -{{d^2}\over {dt^2}} + 4 v^2 t^2 + 4 b^2 + 2 v \sg{2}
$$
We also have the zero-mode operators
$$
\Dslash_0 \equiv i\sg{3}{d\over{dt}},\qquad
H_0 \equiv -{{d^2}\over {dt^2}}.
$$
The propagators are defined as
$$
G(x,y) = \bra{y}H^{-1}\ket{x},\qquad
S(x,y) = -i\bra{y}\Dslash^{-1}\ket{x}\sg{3},\qquad
$$
We find
\eqn\propsg{\eqalign{
G(x,y) &=
\pi^{-1/2} \int ds\, e^{- 4 b^2 s} \sqrt{{{2 v}\over {2\sinh{4 s v}}}}
e^{- \half v (x - y)^2 \coth{2 s v} -\half v (x+y)^2 \tanh{2 s v}} \cr
%%% e^{- v (x^2 + {y}^2) \coth{4 s v} + {{2 v}\over\sinh{4 s v}} x y} \cr
S(x,y) &=
\pi^{-1/2} \int ds\, e^{-4 b^2 s} \sqrt{{{2 v}\over {2\sinh{4 s v}}}}
e^{- \half v (x - y)^2 \coth{2 s v} -\half v (x+y)^2 \tanh{2 s v}} \cr
%%%  e^{-v (x^2 + {y}^2) \coth{2 s v} + {{2 v}\over\sinh{4 s v}} x y} \cr
&\times \big\lbrack
{{v}\over {\sinh{2 sv}}}(x - y) \Id
+{{v}\over {\cosh{2 sv}}}(x + y) \sg{2}
+2 b \sg{1}\cosh{2 sv} + 2 i b\sg{3}\sinh{2 sv}\rbrack \cr
}}
Finally, we define the propagators
$$
G_{\pm} = \bra{y} ( H \pm 4 v)^{-1}\ket{x},\qquad
G_0(x,y) = \bra{y}H_0^{-1}\ket{x},\qquad
S_0(x,y) = -i\bra{y}\Dslash_0^{-1}\ket{x}\sg{3}
$$
and find
\eqn\propz{\eqalign{
G_0(x,y) &= \half |x-y|,\cr
S_0(x,y) &=
\half (\theta(x-y) - \theta(y-x))\Id.\cr
}}

The cubic and quartic interaction vertices give rise to the following
kinds of diagrams:
\item{1.} 
Two cubic vertices joined by three bosonic propagators.
\item{2.}
Two cubic vertices joined by one bosonic and two fermionic propagators.
\item{3.}
Two cubic vertices joined by one bosonic and two ghost propagators.
\item{4.}
A ``figure-of-eight'' bosonic diagram with a single
quartic vertex.
We will evaluate them in section (6).

% --------------------------------------------------------------------- %
% Tadpole diagrams
% --------------------------------------------------------------------- %
\subsec{Diagrams with tadpoles}

The zeroth order approximation of \DKPS\ was taking the D0-branes to
move in a straight line.
At the order of $L$-loops, one will in general find a correction
to the classical trajectory. Thus, the $(L+1)$-th order calculation
should start by  substituting the $L$-th order corrected classical
trajectory. This is tantamount to saying that the $L$-th order
tadpoles cancel and thus we should only include 1PI
diagrams if we expand around the corrected trajectory.
Note that the corrected trajectory can differ by a large distance
from the 0-th order one for times large enough.
In our case, the 1-loop effective potential vanishes up to
order $\Ol{v^3}$, so 
we can just as well keep only the 1PI diagrams at 2-loop order.

% ===================================================================== %
% Section (5): Wave function contribution
% ===================================================================== %
\newsec{Wave-function contribution}

At the order of two-loops the ``profile'' of
the bound-state wave-function could also contribute.
For simplicity, we will restrict our discussion to the original
BFSS model for flat $\MR{11}$.
To see how the contribution arises,
let us write down our Hamiltonian for $U(2N)$ as
$$
{\cal H} = {\cal H}_0 + {\cal H}_{{\rm SU(N)}} 
            + {\cal H}_{{\rm SU(N)}}'
           + V_0 + U
$$
where ${\cal H}_0 + V_0$ is the piece containing only the
$U(1)$-parts of the two $U(N)$-s (i.e. the collective coordinates).
${\cal H}_0$ is the quadratic part that we used in our definition
of the propagators $S(x,y)$ and $G(x,y)$ and $V_0$ are
the interactions.
${\cal H}_{{\rm SU(N)}}$ is the $SU(N)$ part of the Hamiltonian,
which has 16 supersymmetries and is identical to the BFSS Hamiltonian.
${\cal H}_{{\rm SU(N)}}'$ is the $SU(N)$ Hamiltonian for the
other $SU(N)$ corresponding to the second D0-brane bound state.
Finally, $U$ consists of the off-diagonal interactions
which contain interactions between the $(N,\widebar{N})$ fields
and the $SU(N)$ variables.
%%% \eqn\hterms{\eqalign{
%%% {\cal H}_0 &= \cdots,\cr
%%% }}
In all our previous calculations we neglected $U$, and thus,
the pieces ${\cal H}_{{\rm SU(N)}}+{\cal H}_{{\rm SU(N)}}'$ decoupled.

Let us see how $U$ contributes.
Let $X_\u,X_\u'$ be the $U(N)$ coordinates.
we defined
$$
X_\u = \hX_\u  + x_\u\Id,\qquad
X_\u' = \hX_\u'  + x_\u'\Id,
$$
where $\trp{\hX_\u} = 0$.
Let $\psi_\v$ denote the off-diagonal $(N,\widebar{N})$ 
fermions.
Then, $U$ contains terms like:
$$
U = \widebar{\psi}\hX\psi + \cdots
$$
Such a term will give in two-loops an expression like
$$
\langle \widebar{\psi}\psi\widebar{\psi}\psi \rangle
\bra{\Psi_0}\hX\hX\ket{\Psi_0}
$$
(plus a bosonic contribution)
where $\ket{\Psi_0}$ is the ground-state wave-function of $SU(N)$.

This will contribute
\eqn\sswpr{
\int dx\, dy\, (S(x,y) S(x,y)^\dagger  + \px{y}G(x,y) \px{x} G(x,y) + \cdots)
\wpr(x-y)
}
where
$$
\delta_{ab}\wpr(t'-t) \equiv \bra{\Psi_0}T \hX_a (t') \hX_b (t) \ket{\Psi_0}.
$$
We claim that for the order in which we are interested, the 
contribution of $\wpr(x-y)$ is determined by the commutation
relations.
The reason is that $S(x,y)$ in \sswpr\ localizes the $x-y$ variable
to the vicinity of $(x-y)\sim (1/b)\ll 1$.
Thus, we may expand
\eqn\maywp{\eqalign{
\bra{\Psi_0}T \hX_a (t') \hX_b (t) \ket{\Psi_0} &= 
\bra{\Psi_0} \hX_a \hX_b \ket{\Psi_0} + 
\half \bra{\Psi_0} \com{{{d\hX_a}\over {dt}}}{\hX_b} \ket{\Psi_0} |t'-t|
+O(|t'-t|^2),\cr
&= \delta_{ab} (C + \half i |t'-t| +O(|t'-t|^2)),\cr
}}
(We assume here that $\evtr{\hX^2}$ is finite
in the bound state, otherwise the behaviour as a function of 
$t-t'$ might be more singular.)
We do not know the coefficient $C$ which is related to the ``size''
of the bound-state, but when we plug this back to \sswpr\ the total
contribution of $C$ will vanish because the bosonic diagrams will
cancel the fermionic ones. This is as expected, since $C$ will
give a contribution which is larger by a factor of $b$ than the rest
of the diagrams.
The remaining contribution of $|t'-t|$, {\it which was determined solely
from the commutation relations}, is the same as the
free propagator and joins the diagrams with a free propagator, that
we encountered before in such a way that the diagram will scale 
as $N^3$ like all the other diagrams that contribute.

% ===================================================================== %
% Section (6): Wave function contribution
% ===================================================================== %
\newsec{Evaluating the diagrams}

The diagrams with two cubic bosonic vertices give (the $J_k$ integrals
are define in the appendix):
\eqn\allcutog{\eqalign{
\Delta_{{\rm bosonic}}^{(3-3)} &=
80 J_1 + 8 J_3 + 2 J_2 + 64 J_4 + 14 J_5 + 2 J_6 + 4 J_7 - {1\over 2}J_8
-7 J_9 +{7\over 2}J_{11} + {1\over 4}J_{13}
\cr &
+{7\over 2}J_{14} + {9\over 8} J_{15} + 14 J_{16} - 14 J_{17}
+ 3 J_{18} + 2 J_{20}
\cr
&= -{{47\pi}\over {64 b v}} - {{1415\pi v}\over {8192 b^5}}.
\cr
}}
The diagrams with one quartic bosonic vertex give (with 
the definitions of appendix B):
\eqn\allqtg{\eqalign{
\Delta_{{\rm bosonic}}^{(4)} &=
7 K_1 +4 (K_2^{(+)} + K_2^{(-)})
     +{3\over 4}(K_3^{(+)} + K_3^{(-)}) - \half K_4
\cr
&=
+{{\pi}\over {b v}} + {{37 \pi v}\over {512 b^5}}.
\cr
}}
The ghost diagrams give:
\eqn\allgh{
\Delta_{{\rm ghosts}}^{(3-3)} = -4(J_1+J_4) =
  -{{\pi}\over {64 b v}} + {{17 \pi v}\over {8192 b^5}},
}
Thus in total the ``bosonic'' contribution is
\eqn\bostot{
\Delta_{{\rm bosonic}}= \Delta_{{\rm bosonic}}^{(3-3)}+
\Delta_{{\rm bosonic}}^{(4)} + \Delta_{{\rm ghosts}}^{(3-3)} =
{{\pi}\over {4 b v}} - {{403 \pi v}\over {4096 b^5}}}

The fermionic diagrams give 
(the $I$-s are defined in appendix B as well):
\eqn\allf{\eqalign{
%%% \Delta_{{\rm fermionic}}^{(3-3)} &=
\Delta_{{\rm fermionic}} &=
-12 I_2
- (I_6^{(+)} + I_6^{(-)})  
- 2 I_4  
+12 I_5
-2 I_7
- (I_8^{(+)} + I_8^{(-)})
-4 I_3 
-4 I_1
\cr
&= 
-{{\pi}\over {4 b v}} + {{25\pi v}\over {256 b^5}}.
\cr
}}
Altogether we find the phase shift:
\eqn\altogether{
%%% \Delta = \Delta_{{\rm fermionic}}^{(3-3)} + 
%%%  \Delta_{{\rm bosonic}}^{(3-3)} +
%%%  \Delta_{{\rm bosonic}}^{(4)} +
%%%  \Delta_{{\rm ghosts}}^{(3-3)}
\Delta = \Delta_{{\rm fermionic}} + \Delta_{{\rm bosonic}}
 = -{{3\pi v}\over {4096 b^5}}.
}
As expected the terms proportional to ${\pi\over {b v}}$ have
cancelled.

% ===================================================================== %
% Section (8): Discussion
% ===================================================================== %
\newsec{Discussion}

We have found that the 0+1D $U(N)\times U(N)$
Quantum mechanical model for
$\MR{8}/\BZ_2$ predicts the leading phase shift to be
$$
\delta_{{\rm Matrix}}(b,v) = -{{3\pi N^3 v}\over {4096 b^5}}
$$
for the scattering of a bound state of $N$ partons off
the orbifold point (using \NFL).
This result was obtained at two loop order of perturbation theory.

Our system is special in that it exhibits both a non-vanishing 
1-loop contribution which behaves at leading order as 
${{N^2 v^3}\over {b^6}}$
and a non-vanishing 2-loop contribution which behaves at leading order 
as ${{N^3 v}\over {b^5}}$ \NFL.
The 1-loop contribution follows from a 0-brane interacting with
its $\BZ_2$ image to give a ${{N^2 v^4}\over {b^7}}$ potential.
It is intriguing that the two-loop contribution is the dominant one
for small $v$ and large $b$ even at finite $N$.

On the other hand, supergravity predicts a phase shift of (for a charge
$N_2= -2\times {1\over 16}$ in the covering space in \phsf) 
$$
\delta_{{\rm SUGRA}}(b,v) = -{{3\pi N v}\over {16 b^5}}.
$$
Thus we seem to be off by a factor of ${{N^2}\over {256}}$.
It is curious that we get the correct answer for $N=16$,
which might have an interpretation in the context of \SusII.

This is perhaps the simplest example of a discrepancy in a system
with 8 supersymmetries. Such systems have already posed problems
in the past \DOS. In that case the problem arises only when one
blows up the orbifold point. Our example is singular in that
it cannot be blown-up to a smooth CY manifold.
From this point of view, perhaps it is not all that surprising
to find a discrepancy.
Trying to fix this problem promises to teach us something new
about how to compactify M(atrix)-theory on curved manifolds.

Does this mean that the model for $\MR{8}/\BZ_2$ is incorrect?
It seems very hard to add more sectors to the 0+1D quantum
mechanics to correct this result.
We point out that the 0+1D model can be obtained
from a reduction of a chiral model in 1+1D with $(0,8)$
supersymmetry. According to \EBS\ a theory with
1 multiplet containing the $U(N)\times U(N)$ gauge
field and $X_1,X_1'$ fields together with the $\rep{8}_s$ fermionic fields
and 1 multiplet consisting  of  the 8 fields $X_a$ in the
$(N,\bar{N})$ together with the $\rep{8}_c$ fermions
-- is free of anomaly. 

Before concluding we would like to add a caveat.
We have not proven that the our results survive when one takes
$N\rightarrow \infty$ first while keeping $b$ and $P_\perp$
fixed. In principle, the perturbation parameter is $N/b^3$
and the behaviour of M(atrix)-theory could differ from what
we have suggested.\foot{We are grateful to T. Banks for discussions
on the issue of the large $N$ limit in M(atrix)-theory.}

%=======================================================================%
% Acknowledgments
%=======================================================================%
\bigbreak\bigskip\bigskip
\centerline{\bf Acknowledgments}\nobreak
It is a pleasure to thank V. Balasubramanian,
T. Banks, A. Hashimoto, P. Horava, G. Lifschytz, V. Periwal,
S.-J. Rey, E. Silverstein, S. Sethi, S. Shenker,
L. Susskind and O. Tafjord for very helpful
conversations. We would also like to thank all the participants of
the Princeton M(atrix) Journal Club for the discussions.

%=======================================================================%
% Appendix 1:
%=======================================================================%
\bigbreak\bigskip\bigskip
\centerline{\bf Appendix A: The full Lagrangian for $\MR{8}/\BZ_2$.}
\nobreak

With the definitions:
$$
\ImX_2 = {{\Phi_{+} - \Phi_{-}} \over {\sqrt{2}}},\qquad
A_0 - A_0' = \sqrt{2}(\Phi_{+} + \Phi_{-}),
$$
the bosonic quadratic terms in the Lagrangian are:
\eqn\Lbstwox{\eqalign{
L_b^{(2)} &=
(\px{t}\ReX_a)^2 + (\px{t}\ReX_2)^2 + (\px{t}\ReX_3)^2
\cr
&+{1\over 4}\lbrack\px{t}(X_1 + X_1')\rbrack^2
+ {1\over 4}\lbrack\px{t}(A_0 + A_0')\rbrack^2 
\cr
&+(\px{t}\ImX_a)^2  - 4 (b^2 + v^2 t^2) (\ImX_a)^2 
+(\px{t}\ImX_3)^2  - 4 (b^2 + v^2 t^2) (\ImX_3)^2
\cr
&+(\px{t}\Phi_{+})^2  - 4 (b^2 + v^2 t^2 + v) \Phi_{+}^2 
+(\px{t}\Phi_{-})^2  - 4 (b^2 + v^2 t^2 - v) \Phi_{-}^2 
\cr
&+{1\over 4}\lbrack\px{t}(X_1 - X_1')\rbrack^2  
- (b^2 + v^2 t^2) (X_1 - X_1')^2 
\cr
}}
the cubic terms:
%%% \eqn\Lbscub{\eqalign{
%%% L_b^{(3)} &=
%%% -8 b (\ImX_a) (\ImX_3) (\ReX_a) 
%%% - 4 \sqrt{2} b (\ImX_3) \Phi_{+} (\ReX_2) 
%%% + 4 \sqrt{2} b (\ImX_3) \Phi_{-} (\ReX_2) 
%%% \cr &
%%% + 8 b (\ImX_a)^2 (\ReX_3) 
%%% + 8 b \Phi_{+}^2 (\ReX_3) 
%%% + 8 b \Phi_{-}^2 (\ReX_3) 
%%% \cr &
%%% - 4 \sqrt{2} \gamma (\ImX_a) \Phi_{+} (\ReX_a) \tau 
%%% + 4 \sqrt{2} \gamma (\ImX_a) \Phi_{-} (\ReX_a) \tau 
%%% + 8 \gamma (\ImX_a)^2 (\ReX_2) \tau 
%%% \cr &
%%% + 8 \gamma (\ImX_3)^2 (\ReX_2) \tau 
%%% + 4 \gamma \Phi_{+}^2 (\ReX_2) \tau 
%%% + 8 \gamma \Phi_{+} \Phi_{-} (\ReX_2) \tau 
%%% \cr &
%%% + 4 \gamma \Phi_{-}^2 (\ReX_2) \tau 
%%% - 4 \sqrt{2} \gamma (\ImX_3) \Phi_{+} (\ReX_3) \tau 
%%% + 4 \sqrt{2} \gamma (\ImX_3) \Phi_{-} (\ReX_3) \tau 
%%% \cr &
%%% + 2 b (\ReX_3) (X_1 - X_1')^2 
%%% + 2 \gamma (\ReX_2) \tau (X_1 - X_1')^2 
%%% - 2 \sqrt{2} \Phi_{+} (\ReX_3) (\px{\tau}\ImX_3) 
%%% \cr &
%%% - 2 \sqrt{2} \Phi_{-} (\ReX_3) (\px{\tau}\ImX_3) 
%%% + 2 \sqrt{2} \Phi_{+} (\ImX_a) (\px{\tau}\ReX_a) 
%%% - 2 \sqrt{2} \Phi_{+} (\ReX_a) (\px{\tau}\ImX_a) 
%%% \cr &
%%% + 2 \sqrt{2} \Phi_{-} (\ImX_a) (\px{\tau}\ReX_a) 
%%% - 2 \sqrt{2} \Phi_{-} (\ReX_a) (\px{\tau}\ImX_a) 
%%% + 2 \sqrt{2} (\ImX_3) \Phi_{+} (\px{\tau}\ReX_3) 
%%% \cr &
%%% + 2 \sqrt{2} (\ImX_3) \Phi_{-} (\px{\tau}\ReX_3)
%%% + 3 \Phi_{+}^2 (\px{\tau}\ReX_2) 
%%% - 3 \Phi_{-}^2 (\px{\tau}\ReX_2) 
%%% \cr &
%%% - 2 \Phi_{-} (\ReX_2) (\px{\tau}\Phi_{+}) 
%%% + 2 \Phi_{+} (\ReX_2) (\px{\tau}\Phi_{-}) 
%%% \cr
%%% }}
\eqn\Lbscub{\eqalign{
L_b^{(3)} &=
8 b (\ImX_a) (\ImX_3) (\ReX_a) + 8 b (\ImX_2) (\ImX_3) (\ReX_2) 
+ 2 b (A_0 - A_0')^2 (\ReX_3) 
\cr &
- 8 b (\ImX_a)^2 (\ReX_3) - 8 b (\ImX_2)^2 (\ReX_3) 
+ 8 v t (\ImX_a) (\ImX_2) (\ReX_a) 
\cr &
+ 2 v t (A_0 - A_0')^2 (\ReX_2) - 8 v t (\ImX_a)^2 (\ReX_2) 
- 8 v t (\ImX_3)^2 (\ReX_2) 
\cr &
+ 8 v t (\ImX_2) (\ImX_3) (\ReX_3) - 2 b (\ReX_3) (X_1 - X_1')^2 
- 2 v t (\ReX_2) (X_1 - X_1')^2 
\cr &
- 2 (A_0 - A_0') (\ReX_a) (\px{t}\ImX_a) 
- 2 (A_0 - A_0') (\ReX_2) (\px{t}\ImX_2) 
\cr &
- 2 (A_0 - A_0') (\ReX_3) (\px{t}\ImX_3) 
+ 2 (A_0 - A_0') (\ImX_a) (\px{t}\ReX_a) 
\cr &
+ 2 (A_0 - A_0') (\ImX_2) (\px{t}\ReX_2) 
+ 2 (A_0 - A_0') (\ImX_3) (\px{t}\ReX_3)
\cr
}}
and the quartic terms
%%% \eqn\Lbfour{\eqalign{
%%% L_b^{(4)} &=
%%% 2 (\ImX_a)^2 \Phi_{+}^2 + 2 (\ImX_3)^2 \Phi_{+}^2 + \Phi_{+}^4 
%%% + 4 (\ImX_a)^2 \Phi_{+} \Phi_{-} + 4 (\ImX_3)^2 \Phi_{+} \Phi_{-} 
%%% \cr &
%%% + 2 (\ImX_a)^2 \Phi_{-}^2 + 2 (\ImX_3)^2 \Phi_{-}^2 
%%%  - 2 \Phi_{+}^2 \Phi_{-}^2 + \Phi_{-}^4 + 2 (\ImX_b)^2 (\ReX_a)^2 
%%% \cr &
%%% + 4 (\ImX_3)^2 (\ReX_a)^2 + 4 \Phi_{+}^2 (\ReX_a)^2 
%%%   + 4 \Phi_{-}^2 (\ReX_a)^2 - 4 (\ImX_a) (\ImX_b) (\ReX_a) (\ReX_b) 
%%% \cr &
%%% + 2 (\ImX_a)^2 (\ReX_b)^2 - 4 \sqrt{2} (\ImX_a) \Phi_{+} (\ReX_a) (\ReX_2) 
%%% + 4 \sqrt{2} (\ImX_a) \Phi_{-} (\ReX_a) (\ReX_2) 
%%% \cr &
%%%   + 4 (\ImX_a)^2 (\ReX_2)^2 + 4 (\ImX_3)^2 (\ReX_2)^2 
%%% + 2 \Phi_{+}^2 (\ReX_2)^2 + 4 \Phi_{+} \Phi_{-} (\ReX_2)^2 
%%% \cr &
%%%   + 2 \Phi_{-}^2 (\ReX_2)^2 - 8 (\ImX_a) (\ImX_3) (\ReX_a) (\ReX_3) 
%%%  -4 \sqrt{2} (\ImX_3) \Phi_{+} (\ReX_2) (\ReX_3) 
%%% \cr &
%%%   + 4 \sqrt{2} (\ImX_3) \Phi_{-} (\ReX_2) (\ReX_3) 
%%% + 4 (\ImX_a)^2 (\ReX_3)^2 + 4 \Phi_{+}^2 (\ReX_3)^2 
%%% \cr &
%%%   + 4 \Phi_{-}^2 (\ReX_3)^2 + (\ImX_a)^2 (X_1 - X_1')^2 
%%%  +(\ImX_3)^2 (X_1 - X_1')^2 + \half\Phi_{+}^2 (X_1 - X_1')^2
%%% \cr &
%%%   - \Phi_{+} \Phi_{-} (X_1 - X_1')^2 
%%%  +\half\Phi_{-}^2 (X_1 - X_1')^2 + (\ReX_a)^2 (X_1 - X_1')^2 
%%% \cr &
%%%   + (\ReX_2)^2 (X_1 - X_1')^2 + (\ReX_3)^2 (X_1 - X_1')^2
%%% \cr
%%% }}
\eqn\Lbfour{\eqalign{
L_b^{(4)} &=
(A_0 - A_0')^2 (\ImX_a)^2 
+ (A_0 - A_0')^2 (\ImX_2)^2 
+ (A_0 - A_0')^2 (\ImX_3)^2 
\cr &
+ (A_0 - A_0')^2 (\ReX_a)^2 
- 2 (\ImX_b)^2 (\ReX_a)^2 
- 4 (\ImX_2)^2 (\ReX_a)^2 
\cr &
- 4 (\ImX_3)^2 (\ReX_a)^2 
+ 4 (\ImX_a) (\ImX_b) (\ReX_a) (\ReX_b) 
- 2 (\ImX_a)^2 (\ReX_b)^2 
\cr &
+ 8 (\ImX_a) (\ImX_2) (\ReX_a) (\ReX_2) 
+ (A_0 - A_0')^2 (\ReX_2)^2 
- 4 (\ImX_a)^2 (\ReX_2)^2 
\cr &
- 4 (\ImX_3)^2 (\ReX_2)^2 
+ 8 (\ImX_a) (\ImX_3) (\ReX_a) (\ReX_3) 
\cr &
+ 8 (\ImX_2) (\ImX_3) (\ReX_2) (\ReX_3) 
+ (A_0 - A_0')^2 (\ReX_3)^2 
\cr &
- 4 (\ImX_a)^2 (\ReX_3)^2 
- 4 (\ImX_2)^2 (\ReX_3)^2 
- (\ImX_a)^2 (X_1 - X_1')^2 
\cr &
- (\ImX_2)^2 (X_1 - X_1')^2 
- (\ImX_3)^2 (X_1 - X_1')^2 
\cr &
- (\ReX_a)^2 (X_1 - X_1')^2 
- (\ReX_2)^2 (X_1 - X_1')^2 
\cr &
- (\ReX_3)^2 (X_1 - X_1')^2
\cr
}}
The fermionic variables appear in:
\eqn\fermlg{\eqalign{
L_f &=
i\xi_i^\dagger \px{t}\xi^i + i\eta_i^\dagger \px{t}\eta^i
+2 v t \eta_i^\dagger\sg{2}\eta^i + 2 b \eta_i^\dagger\sg{1}\eta^i \cr
%%% &-{i\over 2}(X^a + {X^a}^\dagger)
&-i(\ReX_a)
\{{\eta^i}^T\sg{1}\eta^j\gamma_{ij}^a +
  {\bret_i}^T\sg{1}{\bret_j}\gamma^{ij,a}\} \cr
%%% &+{1\over 2} (X^2 - {X^2}^\dagger) 
&+ {i\over {\sqrt{2}}}(\Phi_{-} - \Phi_{+})
   \{ \eta_i^\dagger\sg{+}\xi^i - {\xi^i}^\dagger\sg{-}\eta_i \} 
%%% +{i\over 2} (X^3 - {X^3}^\dagger) 
-(\ImX_3)
   \{ \eta_i^\dagger\sg{+}\xi^i + {\xi^i}^\dagger\sg{-}\eta_i \}  \cr
%%% &+{1\over 2} (X^a - {X^a}^\dagger) 
&+i(\ImX_a)
   \{ {\eta^i}^T\sg{+}\xi^j\gamma^a_{ij}  
      + {\xi_i}^\dagger\sg{-}\bret_j\gamma^{ij,a} \} 
+ {i\over 2}(X^1 - {X^1}')
   \{\xi_i^\dagger (\Id - \sg{3})\eta^i
    - \eta_i^\dagger (\Id-\sg{3})\xi^i \} \cr
%%% &+ {i\over 2}(A_0 - A_0')
&+ {{\sqrt{2} }\over 2}i(\Phi_{-} + \Phi_{+})
  \{\xi_i^\dagger (\Id - \sg{3})\eta^i
    - \eta_i^\dagger (\Id-\sg{3})\xi^i \} \cr
%%% &+(X_2 + X_2^\dagger)
&+2(\ReX_2)
\eta_i^\dagger\sg{2}\eta^i
%%% +(X_3 + X_3^\dagger)
+2(\ReX_3)
\eta_i^\dagger\sg{1}\eta^i\cr
}}

%=======================================================================%
% Appendix 2:
%=======================================================================%
\bigbreak\bigskip\bigskip
\centerline{\bf Appendix B: Integrals.}
\nobreak

The fields have the following propagators:
\eqn\allpr{\eqalign{
\ReX_2 &\rightarrow  {1\over {\sqrt{2}}} G_0,\cr
\ImX_3 &\rightarrow  {1\over {\sqrt{2}}} G,\cr
\ReX_3 &\rightarrow  {1\over {\sqrt{2}}} G_0,\cr
\ImX_a &\rightarrow  {1\over {\sqrt{2}}} G,\cr
\ReX_a &\rightarrow  {1\over {\sqrt{2}}} G_0,\cr
\Phi_{-}   &\rightarrow  {1\over {\sqrt{2}}} G_{-},\cr
\Phi_{+}   &\rightarrow  {1\over {\sqrt{2}}} G_{+},\cr
A_0 + A_0'   &\rightarrow  {\sqrt{2}} G_0,\cr
X_1 - X_1'   &\rightarrow  {\sqrt{2}} G,\cr
\chi         &\rightarrow S_0,\cr
\eta         &\rightarrow S,\cr
}}

%%% We have define
%%% $$
%%% \ImX_2 = {{\Phi_{+} - \Phi_{-}} \over {\sqrt{2}}},\qquad
%%% A_0 - A_0' = \sqrt{2}(\Phi_{-} + \Phi_{+})
%%% $$

%------------------------------------------------------------------%
% Cubic terms
%------------------------------------------------------------------%
We are going to need a few integrals.
We calculated them with the definitions:
\eqn\propdf{\eqalign{
G(x,y) &=
\pi^{-1/2} \int ds\, e^{- 4 b^2 s} \sqrt{{{2 v}\over {2\sinh{4 s v}}}}
e^{- \half v (x - y)^2 \coth{2 s v} -\half v (x+y)^2 \tanh{2 s v}} \cr
S(x,y) &=
\pi^{-1/2} \int ds\, e^{-4 b^2 s} \sqrt{{{2 v}\over {2\sinh{4 s v}}}}
e^{- \half v (x - y)^2 \coth{2 s v} -\half v (x+y)^2 \tanh{2 s v}} \cr
&\times \big\lbrack
{{v}\over {\sinh{2 sv}}}(x - y) \Id
+{{v}\over {\cosh{2 sv}}}(x + y) \sg{2}
+2 b \sg{1}\cosh{2 sv} + 2i b\sg{3}\sinh{2 sv}\rbrack \cr
G_0(x,y) &= \half |x-y|,\cr
S_0(x,y) &=
\half (\theta(x-y) - \theta(y-x))\Id,\cr
G_{\pm}(x,y) &= 
\pi^{-1/2} \int ds\, e^{\pm 4 v s - 4 b^2 s} \sqrt{{{2 v}\over {2\sinh{4 s v}}}}
e^{- \half v (x - y)^2 \coth{2 s v} -\half v (x+y)^2 \tanh{2 s v}} \cr
S_{\pm}(x,y) &=
\pi^{-1/2} \int ds\, e^{\pm 4 v s -4 b^2 s} \sqrt{{{2 v}\over {2\sinh{4 s v}}}}
e^{- \half v (x - y)^2 \coth{2 s v} -\half v (x+y)^2 \tanh{2 s v}}. \cr
}}
For the bosonic diagrams with two cubic vertices we will need:
\eqn\ggg{\eqalign{
J_1 = &
b^2\int dx\, dy\, G_0 G G 
  = {{\pi}\over{512 b v}} -{{95\pi v}\over{262144 b^5}} + \Ol{v^2},
\cr
J_2 = &
b^2\int dx\, dy\, G_0 G (G_{+} + G_{-})
  = {{\pi}\over{256 b v}} +{{205\pi v}\over{131072 b^5}} + \Ol{v^2},
\cr
J_3 = &
b^2\int dx\, dy\, G_0 (G_{+} G_{+} + G_{-} G_{-})
  = {{\pi}\over{256 b v}} +{{865\pi v}\over{131072 b^5}} + \Ol{v^2},
\cr
}}

\eqn\xyggg{\eqalign{
J_4 = &
v^2\int dx\, dy\,\,\, x y G_0 G G
  = {{\pi}\over{512 b v}} -{{41\pi v}\over{262144 b^5}} + \Ol{v^2},
\cr
J_{5} = &
v^2\int dx\, dy\, x y\, G_0 G (G_{+} + G_{-})
 =  {{\pi}\over{256 b v}} +{{19\pi v}\over{131072 b^5}} + \Ol{v^2},
\cr
J_6 = &
v^2\int dx\, dy\,\,\, x y G_0 (G_{+} G_{+} + G_{-} G_{-})
   = {{\pi}\over{256 b v}} +{{151\pi v}\over{131072 b^5}} + \Ol{v^2},
\cr
J_7 = &
v^2\int dx\, dy\,\,\, x y G_0 G_{+} G_{-}
  = {{\pi}\over{512 b v}} +{{7\pi v}\over{262144 b^5}} + \Ol{v^2},
\cr
}}

\eqn\ggpxgpy{\eqalign{
J_{8} = &
\int dx\, dy\, G_0\px{x}G_{+}\px{y}G_{-}
 =  -{{\pi}\over{64 b v}} -{{17\pi v}\over{16384 b^5}} + \Ol{v^2},
\cr
J_{9} = &
\int dx\, dy\, (G_{+} + G_{-})\px{y}G\px{x}G_0
 = {{\pi}\over{16 b v}} +{{53\pi v}\over{4096 b^5}} + \Ol{v^2},
\cr
J_{10} = &
\int dx\, dy\, (G_{+}\px{x}G_{+} +G_{-}\px{x}G_{-})\px{y}G_0
 =  {{\pi}\over{16 b v}} +{{89\pi v}\over{4096 b^5}} + \Ol{v^2},
\cr
}}

\eqn\ggpxpyg{\eqalign{
J_{11} = &
\int dx\, dy\, G_0(G_{+} + G_{-})\px{x}\px{y}G
 =  -{{\pi}\over{32 b v}} -{{83\pi v}\over{8192 b^5}} + \Ol{v^2},
\cr
J_{12} = &
\int dx\, dy\, G_0 (G_{+}\px{x}\px{y}G_{+} + G_{-}\px{x}\px{y}G_{-}) 
  = -{{\pi}\over{32 b v}} -{{89\pi v}\over{8192 b^5}} + \Ol{v^2},
\cr
J_{13} = &
\int dx\, dy\, G_0 (G_{-}\px{x}\px{y}G_{+} + G_{+}\px{x}\px{y}G_{-})
 =  -{{\pi}\over{32 b v}} -{{65\pi v}\over{8192 b^5}} + \Ol{v^2},
\cr
J_{14} = &
\int dx\, dy\, (G_{+} + G_{-}) G \px{x}\px{y}G_0
  = -{{\pi}\over{8 b v}} -{{29\pi v}\over{2048 b^5}} + \Ol{v^2},
\cr
J_{15} = &
\int dx\, dy\, (G_{+}G_{+}+ G_{-}G_{-})\px{x}\px{y}G_0
   = -{{\pi}\over{8 b v}} -{{89\pi v}\over{2048 b^5}} + \Ol{v^2},
\cr
}}

\eqn\vxggpg{\eqalign{
J_{16} = &
v \int dx\, dy\, x\, G_0 (G_{+} - G_{-})\px{y}G
 =  0 +{{21\pi v}\over{16384 b^5}} + \Ol{v^2},
\cr
J_{17} = &
v \int dx\, dy\, y\, G (G_{+} - G_{-})\px{x}G_0
 =  0 -{{3\pi v}\over{2048 b^5}} + \Ol{v^2},
\cr
J_{18} = &
v \int dx\, dy\, y\, (G_{+}G_{+} - G_{-}G_{-})\px{x}G_0
 =  0 -{{3\pi v}\over{1024 b^5}} + \Ol{v^2},
\cr
J_{19} = &
v \int dx\, dy\, y\, G_0(G_{+}\px{x}G_{+} - G_{-}\px{x}G_{-})
 =  0 +{{3\pi v}\over{2048 b^5}} + \Ol{v^2},
\cr
J_{20} = &
v\int dx\, dy\, y\, G_0 (G_{+}\px{x}G_{-} - G_{-}\px{x}G_{+})
 =  0 + {{9\pi v}\over {8192 b^5}} + \Ol{v^2},
\cr
}}

For the ``figure of eight'' diagrams with one quartic vertex 
we will need:
\eqn\Ks{\eqalign{
K_1 &=
\int dx\, G(x,x) G(x,x) dx  =
{{\pi}\over{16 b v}} -{{7\pi v}\over{4096 b^5}} + \Ol{v^2},\cr
K_2^{(\pm)} &=
\int dx\, G(x,x) G_{+}(x,x) dx  =
\pm {\pi\over{64 b^3}} +
{{\pi}\over{16 b v}} +{{29\pi v}\over{4096 b^5}} + \Ol{v^2},\cr
K_3^{(\pm)} &=
\int dx\, G_{+}(x,x) G_{+}(x,x) dx  =
\pm {\pi\over{32 b^3}} +
{{\pi}\over{16 b v}} +{{89\pi v}\over{4096 b^5}} + \Ol{v^2},\cr
K_4 &=
\int dx\, G_{+}(x,x) G_{-}(x,x) dx  =
{{\pi}\over{16 b v}} +{{41\pi v}\over{4096 b^5}} + \Ol{v^2}.\cr
}}

and for the diagrams with two cubic fermionic vertices we need:
\eqn\Is{\eqalign{
I_1 & =
\int dx\, dy\,
\trp{S(x,y) \sg{1} S(y,x)\sg{1}} G_{0}(x,y)
 =  -{{\pi}\over{32 b v}} +{{125\pi v}\over{16384 b^5}} + \Ol{v^2},
\cr
I_2 & =
\int dx\, dy\,
\trp{S(x,y) \sg{1} S(y,x)^* \sg{1}} G_{0}(x,y)
 =  0,
\cr
I_3 & =
\int dx\, dy\,
\trp{S(x,y) \sg{2} S(y,x) \sg{2}} G_{0}(x,y)
 =  -{{\pi}\over{32 b v}} +{{35\pi v}\over{16384 b^5}} + \Ol{v^2},
\cr
I_4 & = 
\int dx\, dy\,
\trp{S_0(x,y) \sg{-} S(y,x) \sg{+}} G(x,y)
 = 
-{{\pi}\over{8 b v}} +{{5\pi v}\over{1024 b^5}} + \Ol{v^2},
\cr
I_5 & = 
\int dx\, dy\,
\trp{S_0(x,y) \sg{-} S(y,x)^* \sg{+}} G(x,y)
  = 
-{{\pi}\over{8 b v}} +{{5\pi v}\over{1024 b^5}} + \Ol{v^2},
\cr
I_6^{(\pm)} & = 
\int dx\, dy\,
\trp{S_0(x,y) \sg{-} S(y,x) \sg{+}} G_{\pm}(x,y)
 =
\mp {{3\pi}\over {32 b^3}}
  -{{\pi}\over{8 b v}} -{{25\pi v}\over{1024 b^5}} + \Ol{v^2},
\cr
I_7 & = 
\int dx\, dy\,
\trp{S_0(x,y) (\Id-\sg{3}) S(y,x) (\Id-\sg{3})} G(x,y)
 =
-{{\pi}\over{8 b v}} +{{5\pi v}\over{1024 b^5}} + \Ol{v^2},
\cr
I_8^{(\pm)} & =
\int dx\, dy\,
\trp{S_0(x,y) (\Id-\sg{3})S(y,x) (\Id-\sg{3})} G_{\pm}(x,y)\cr
 &=
\mp {{3\pi}\over {32 b^3}}
  -{{\pi}\over{8 b v}} -{{25\pi v}\over{1024 b^5}} + \Ol{v^2}.
\cr
}}

%------------------------------------------------------------------%
% File name
%------------------------------------------------------------------%
%%% \newsec{Comments}
%%% 
%%% The file name is:
%%% {\bf Paper/paper.tex}

\listrefs
\end